\documentclass[11pt]{article}
\usepackage{graphicx}
\usepackage{amssymb}
\def\lsim{\mathrel{\raise.3ex\hbox{$<$\kern-.75em\lower1ex\hbox{$\sim$}}}}
\def\gsim{\mathrel{\raise.3ex\hbox{$>$\kern-.75em\lower1ex\hbox{$\sim$}}}}

\newcommand{\nc}{\newcommand}
\newcommand{\eq}{Eq.}
\nc{\beq}{\begin{equation}}   \nc{\eeq}{\end{equation}}
\nc{\bea}{\begin{eqnarray}}   \nc{\eea}{\end{eqnarray}}
\nc{\baa}{\begin{array}}      \nc{\eaa}{\end{array}}
\nc{\bit}{\begin{itemize}}    \nc{\eit}{\end{itemize}} 
\nc{\ben}{\begin{enumerate}}  \nc{\een}{\end{enumerate}}
\nc{\bce}{\begin{center}}     \nc{\ece}{\end{center}}
\def\beqa{\begin{eqnarray}}
\def\eeqa{\end{eqnarray}}

\def\ie{{\it i.e.,}}

\begin{document}

\title{Ultra high energy neutrino-nucleon cross section
from \\ cosmic ray experiments and neutrino telescopes}
\author{V. Barger$^1$, Patrick Huber$^1$ and Danny Marfatia$^{2}$\\[2ex]
\small\it $^1$Department of Physics, University of Wisconsin, Madison, WI 53706\\
\small\it $^2$Department of Physics and Astronomy, University of Kansas, Lawrence, KS 66045}

\date{}

\maketitle

\begin{abstract}
We deduce the cosmogenic neutrino flux by jointly analysing 
ultra high energy cosmic ray data 
from HiRes-I and II, AGASA and the Pierre Auger Observatory.  
We make two determinations of the neutrino flux by using
 a model-dependent method and a model-independent method.
The former is well-known, and involves the use
of a power-law injection spectrum. The latter 
is a regularized unfolding procedure.
We then use neutrino flux bounds obtained by the RICE experiment
to constrain the neutrino-nucleon inelastic cross section at energies 
inaccessible at colliders. The cross section bounds obtained using
the cosmogenic fluxes derived by unfolding are the most model-independent
bounds to date.

\end{abstract}

\newpage

\section{Introduction}

Neutrino cross sections at high energies could herald new physics because
several extensions of the Standard Model predict enhanced cross sections. 
Examples include electroweak instanton processes~\cite{instanton}, black hole 
production~\cite{banks} and exchange of towers of 
Kaluza-Klein gravitons~\cite{gravitons} in low 
scale gravity models~\cite{add}, and TeV-scale string 
excitations~\cite{strings}. The highest center-of-mass 
energy at which the neutrino-nucleon 
cross section has been measured is about $300$ GeV at the HERA accelerator.
The only practical way to probe this cross section 
at center-of-mass energies above 100 TeV is by detecting the interactions 
of neutrinos
with energy above $10^{10}$ GeV incident on Earth. By {\it ultra high
energy} one often means energies above $10^9$ GeV, 
but this usage is not standard.

While neutrinos with energies above $\cal{O}$($10^4$)~GeV
have not been observed so far~\cite{Ribordy:2005fi}, they
are expected to accompany ultra high energy cosmic rays (that have
been observed with energies exceeding $10^{10}$ GeV) 
because almost all potential 
cosmic ray sources are predicted to produce protons, neutrinos and gamma rays
with comparable rates. Experiments
indicate that the highest energy cosmic rays are primarily 
protons~\cite{bird}. 
A guaranteed source of neutrinos, the cosmogenic neutrinos, arises from the 
inelastic interactions of cosmic ray protons on the cosmic microwave 
background~\cite{gzk}, prominently $p\gamma \rightarrow \Delta^+ \rightarrow n \pi^+$, 
followed by pion decay~\cite{cosmicnu}.  
The threshold energy for this reaction is the
Gresein-Zatsepin-Kuzmin (GZK) energy, $E_{GZK} \sim 4\times 10^{10}$ 
GeV~\cite{gzk}. 
At such high 
energies, the gyroradius of a proton in the galactic magnetic field is larger
than the size of the Galaxy and it is therefore expected that these cosmic
ray protons are of extragalactic origin. Since the attenuation caused
by the above reaction has a length scale of about 
50 Mpc~\cite{stecker,yoshida}, 
a strong suppression in the cosmic ray 
spectrum is expected above $E_{GZK}$. There is evidence for the GZK cutoff in 
Fly's Eye/HiRes~\cite{hires,berg2} data but not in AGASA~\cite{agasa,takeda} 
data. 
The Pierre Auger Observatory~\cite{pao} 
(which we will 
refer to as Auger in what follows)
is expected to resolve this conflict.

Estimates of the cosmogenic flux are very model-dependent and 
differ by about $2-3$ 
orders of magnitude~\cite{yoshida,cosmogenic,cosmomodels}. 
As we discuss below, the uncertainty in the cosmogenic
flux is further exacerbated by recent evidence from the HiRes experiment  
that protons dominate the
flux at about $5\times 10^{8}$ GeV~\cite{morehires}, 
two orders of magnitude below $E_{GZK}$.

The uncertainty in the cosmogenic neutrino flux must be dealt with before 
progress can be made to extract or constrain the neutrino-nucleon inelastic 
cross section 
using cosmic neutrinos. We prefer to use cosmic ray data
to infer the cosmogenic neutrino flux, thus side-stepping theoretical 
modelling.
So long as a lower
bound on the cosmogenic flux is established, it will be possible
to place an upper bound on the neutrino-nucleon inelastic 
cross section.

In Refs~\cite{ring,ahlers}, 
estimates of the cosmogenic neutrino flux have been made from 
separate analyses of AGASA and HiRes-II data combined with their predecessor
collaborations, Akeno and Fly's Eye, respectively. These results have
been used to constrain the neutrino-nucleon cross section in Ref.~\cite{ring2}.
Our approach is different.

It has been shown in previous work that the apparent disagreement between 
AGASA and HiRes data could be a result of large systematic uncertainties in
the energy determinations of the two experiments~\cite{olinto}. 
The energy scale uncertainty
of the AGASA experiment is 30\%~\cite{takeda}, while that of the HiRes detectors is about
20\% above $10^{10.5}$ GeV and about 25\% above $10^{9.5}$ GeV~\cite{berg}. 
The data
from Auger have an energy uncertainty of about 25\%~\cite{sommers}. 

Rather than perform separate analyses of these datasets, we take into
account the
large uncertainties in the energy measurements and perform a combined analysis
of Auger~\cite{sommers}, HiRes-I (sample collected between June 1997 and 
May 2005) and II 
(sample collected between December 1999 and May 2003)~\cite{berg2}, 
and AGASA data~\cite{agasa}. We expect the double-counting 
of events that are common to the
HiRes-I and HiRes-II datasets to have a negligible effect on our results. 
We suppose that the energy 
uncertainties for HiRes-I and II are the same and allow a total of 3 
floating energy scale parameters: one each for Auger, HiRes and AGASA.

It was thought that the ankle at $10^{10}$ GeV indicates the onset of
the dominance of a 
higher energy flux of particles over a lower energy flux, such as an
extragalactic component starting to dominate over the galactic component
of the cosmic ray flux. 
The dip structure in the vicinity of the ankle 
can be explained as 
a consequence of pair production from extragalactic protons on the 
CMB~\cite{ber}. 
The latter process has a threshold of $10^{8.6}$ GeV thus permitting 
the interpretation that protons dominate even below this energy.
Recent evidence for a steepening of the spectrum at about the same 
energy~\cite{moreagasa}, the
second knee, suggests that
a low energy transition is consistent with the data. 
Moreover, the HiRes collaboration finds a drastic change of 
composition across
the second knee, from about 50\% protons just below this knee to about 80\%
protons above~\cite{morehires}. Since the end-point of 
the galactic flux is thought to be comprised of heavy nuclei, a 
changeover to protons is interpreted as the onset of the dominance of the 
extragalactic flux. We do not analyse events with energy below the second 
knee since they are probably not of cosmological origin. However,
since composition measurements are strongly model-dependent (via the
theoretical uncertainty in predicting electron and muon shower sizes 
and the depth of shower maximum for hadronic showers), the only
statement about the transition energy that can be made with some confidence
is that it lies in the interval $10^{8}$ GeV to $10^{10}$ GeV.

\section{Modelling}

We assume that all observed cosmic ray events above the second knee are due to
protons and that cosmic ray sources are isotropically distributed.
We follow Ref.~\cite{ring} 
and write the differential flux of particles of type $b$ 
($b$ may be protons, $p$, or cosmogenic neutrinos, $\nu_{l,\bar{l}}$), 
  with energy 
$E$ arriving at earth as
\begin{eqnarray}
J_b(E) &\equiv &{\frac{d^4 N_b}{dE\,dA\, dt\, d\Omega}}
       = \int_0^\infty dE_i G_{b}(E,E_i)I(E_i,t) \,,
\label{flux}
\end{eqnarray}
where
\begin{equation}
    G_{b}(E,E_i) = {1\over{4\pi}}
\int_0^\infty dr \left| {\frac{\partial P_{b}(E;E_i,r)}{\partial E}} \right| \rho_{0} [1+z(r)]^n \Theta(z-z_{min}) \Theta(z_{max}-z)\,,
\end{equation} 
\begin{equation}
I(E_i,t)={\frac{d^2{N}_p} {dE_i\, dt}}\,,
\end{equation}
 $N_b$ is the number of $b$ particles and 
$A$, $t$ and $\Omega$ denote area, time and solid angle, respectively.
This equation needs some explanation. $P_{b}(E;E_i,r)$ is the probability 
of a $b$ particle arriving at earth with energy above $E$
if a proton with energy $E_i$ was emitted from a source at distance 
$r$~\cite{propagation}. These
propagation functions are available from Ref.~\cite{prop2}. 
Here, $\rho_0$ is the comoving number density of sources.
The step functions restrict the sources to have redshifts between $z_{min}$
and $z_{max}$. We take $z_{min}=0.012$ (corresponding to 50 Mpc) 
and $z_{max}=2$. 
Redshift evolution of the sources is parameterized by $n$, which
also mimics changes in $z_{max}$. Reference~\cite{ahlers}
has emphasized that since the transition energy is lower than the threshold
for $p\gamma_{CMB}$ absorption,
sources at high redshift are also sampled. Thus, source evolution must
be accounted for.
For pure redshifting, $n=3$. Since $dz=(1+z) H(z) dr$ (where
$H(z)$ is the Hubble parameter), $J_b$ depends on the cosmology describing 
our universe. We adopt a flat cosmological constant-dominated universe
with $\Omega_m=0.3$ and $H(0)=71$ km/s/Mpc, since these values were used
to calculate the propagation functions; our results are insensitive to
 variations in the cosmological model chosen.
$I(E_i,t)$ is the injection spectrum.

Throughout, we assume that extragalactic magnetic fields are smaller 
than $10^{-9}$ G
and therefore neglect synchroton radiation of protons.

We employ two methods for obtaining the proton injection spectrum (which we 
assume to be identical for all sources).

\subsection{Power-law injection spectrum}

In the first method, we adopt 
a power-law 
spectrum with index $\alpha$,
\begin{equation}
I(E_i,t) = I_0 E_i^{\alpha} \Theta(E_{max}-E_i)\,,
\label{eq:powerlaw}
\end{equation}
where $I_0$ the normalizes the injection spectrum and $E_{max}$ is the 
maximum injection energy achievable through astrophysical processes which we
set equal to $3\times 10^{12}$ GeV.
The overall normalization $\rho_0 I_0$ is determined from cosmic
ray data. 

After extracting $\alpha$ from cosmic ray data, 
we can determine the cosmogenic neutrino 
flux resulting from the GZK mechanism by setting $b=\nu$ in Eq.~(\ref{flux}).
(We either fix $n$ or determine it simultaneously with $\alpha$).

For a description of the statistical procedure see Appendix A.

\subsection{Unfolding the injection spectrum}

The essential idea behind deconvolving the injection spectrum from cosmic ray
data is described in Ref.~\cite{ring}. 

The proton
injection spectrum (with statistical uncertainties) 
is obtained by inverting the observed proton spectrum. Use must be made of the 
propagation
function of the proton. Although this function is not invertible in general,
the injection spectrum may be unfolded 
under the assumptions made earlier that
cosmic sources are isotropically distributed 
(within a redshift range, $[z_{min}, z_{max}]$) and 
the redshift evolution 
can be parameterized by $(1+z)^n$. Then, Eq.~(\ref{flux}) takes the form
of a matrix equation after writing the integral as a sum over energy,
\begin{equation}
\bold{I} = \bold{G}_{b}^{-1}\, \bold{J}_b\,,
\end{equation}
where $\bold{I}$ is the discrete version of the injection spectrum 
$I(E_i,t)$. Since cosmic ray data give $\bold{J}_b$ 
with $b=p$, we can find $\bold{I}$.
This seemingly simple procedure is fraught with technical difficulties.
We relegate the details of the statistical methodology to Appendix B. 

It is noteworthy that this method is applicable 
even for AGASA data that do not 
show evidence for
a GZK suppression; events beyond the GZK suppression 
can not be accounted for by a  
single power-law injection spectrum.

The cosmogenic neutrino spectrum (with statistical
uncertainties) is inferred
from the proton spectrum by using
\begin{equation}
\bold{J}_\nu = \bold{G}_{\nu}\, \bold{I}\,.
\label{neutinv}
\end{equation}

A lower bound on the neutrino flux is obtained by assuming that only the
events at and above the pile-up below the GZK energy are protons. This
is consistent with the GZK interpretation that the energy of 
protons close to the GZK energy degrades significantly causing the pile-up.
The upper
bound is found by supposing that all the observed events above the second knee 
are protons. The events above the GZK energy could arise simply because
the
injection spectrum was sufficiently large.

\section{The cosmic ray spectrum and the cosmogenic neutrino flux}

We first consider the case of a power-law injection spectrum. In our analysis
we only consider data in the energy range $10^{9.6}$ GeV to $10^{11}$ GeV.
Including data outside this range gives results which have a
 goodness of fit (gof) below $\cal{O}$$(10^{-10})$. That this is the case
for data above $10^{11}$ GeV is understood as a consequence of the inability
of a power-law injection spectrum to explain super-GZK events. The AGASA data
point at $10^{9.55}$ eV is statistically significant and discrepant with 
data from other experiments.

\begin{figure}[ht]
\centering\leavevmode
\mbox{\includegraphics[width=6.5in]{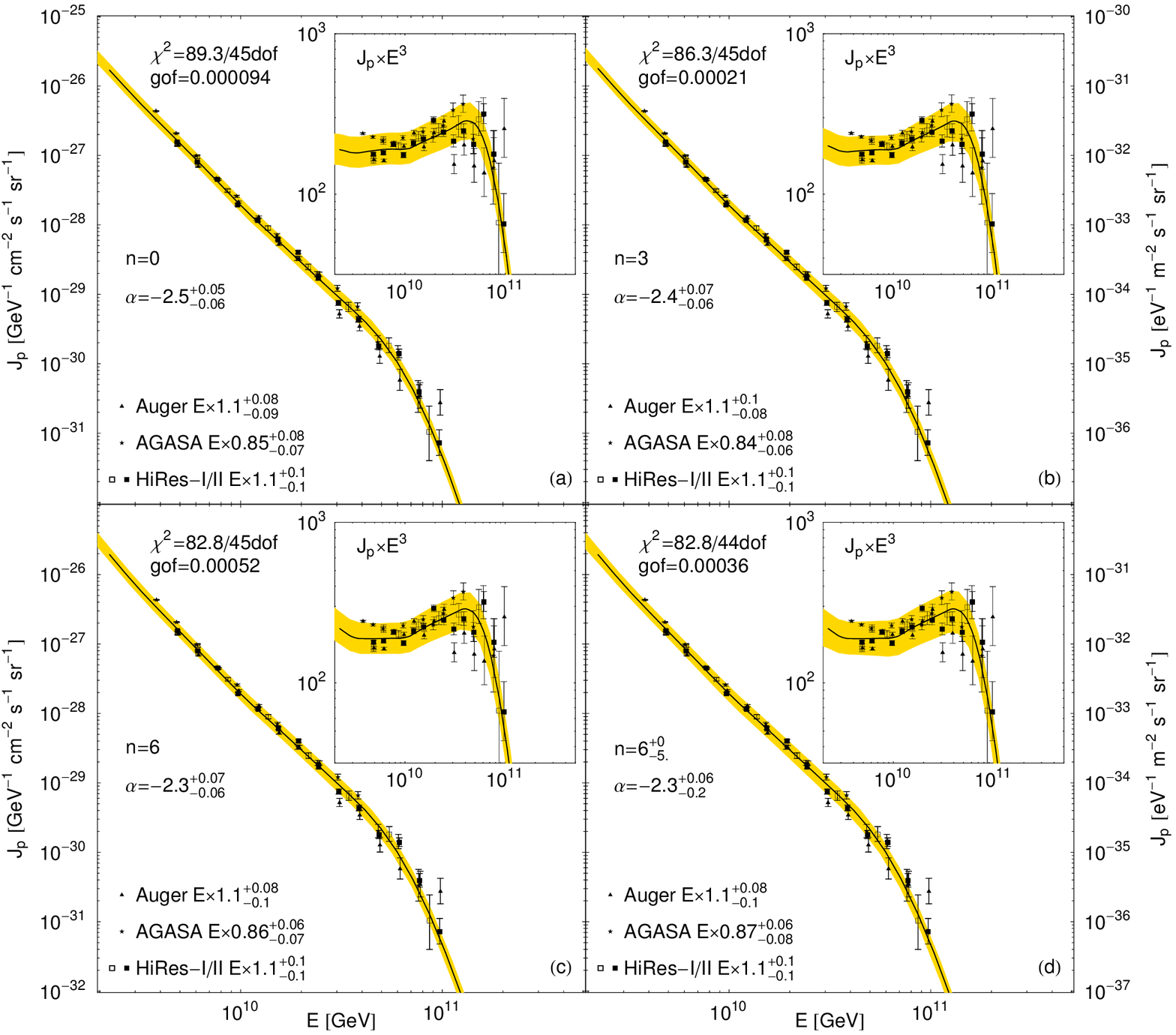}}
\caption[]{The flux of $J_p$ of ultra high energy cosmic rays 
that simultaneously 
fits Auger, HiRes-I and II, and AGASA data in the energy range 
$10^{9.6}$ GeV to $10^{11}$ GeV assuming a power-law injection spectrum 
with spectral index $\alpha$.
Redshift evolution of the sources is parameterized by a power-law,
$(1+z)^n$. The shaded bands and the quoted uncertainties 
are at the 2$\sigma$ C.~L. and the error bars on the data points
are 1$\sigma$ uncertainties. The energy scale of each experiment has been
allowed to vary within the experiment's energy scale uncertainty. The insets
show $J_p E^3$ to facilitate comparison with other analyses.
\label{fig:protons}}
\end{figure}

In Fig.~\ref{fig:protons}, 
we plot $J_p$ vs. $E$ for three different values of $n$ and for the case in
which $n$ is a free parameter in the fit that is allowed to vary between 
0 and 6. The solid line is the best-fit
to the data in each case. In all cases, a spectral index close to $-2.4$ is 
favored. The shaded band and quoted uncertainties
correspond to models that are consistent with
the data at the 2$\sigma$ C.~L. The error bars on the data points
are 1$\sigma$ uncertainties. The insets show $J_p E^3$ vs $E$ to
enable comparison with results presented in other papers. The y-axis units on
the left-hand side are  \mbox{GeV$^{-1}$cm$^{-2}$s$^{-1}$sr$^{-1}$} 
and those on the
right-hand side are \mbox{eV$^{-1}$m$^{-2}$s$^{-1}$ sr$^{-1}$}. Note that since
the energy scales of the experiments have large uncertainties, 
it is misleading to plot $J_p E^3$. From panel (d) it can be
seen that the 2$\sigma$ uncertainty in $n$ spans almost 
the entire range within
which $n$ was permitted to vary. Although $n<1$ is disfavored at 2$\sigma$,
we will show results for $n=0$ because its exclusion is not very significant.

The goodness of fit of these joint analyses is poor. This
is primarily because the initial Auger data are noisy; see Tables~1 and 2. 
For now, we proceed to
determine the corresponding cosmogenic neutrino fluxes.

\begin{table}[ht]
\begin{center}
\label{tab1}
\begin{tabular}{|l|c|c|c|}
\hline
      & $\chi^2$& dof & gof \\ \hline
All data & 86.3 & 45 & $2.1\times 10^{-4}$\\
$-$Auger & 37.6    & 32  & 0.23\\
$-$AGASA & 58.8    & 32  & $2.7\times 10^{-3}$\\
$-$HiRes & 65.1 & 24 & $1.2\times 10^{-5}$\\\hline
\end{tabular}
\end{center}
\caption{The $\chi^2$ for the number of degrees of freedom (dof) and the
corresponding goodness of fit (gof) for analyses with $n=3$. The first row
is for an analysis of all data. The cosmic ray spectrum is shown in the
panel (b) of Fig.~\ref{fig:protons}. Each subsequent row is the result
with the indicated dataset removed from the analysis. Removal of the Auger
data
from the analysis improves the gof considerably.}
\end{table}

\begin{table}[ht]
\begin{center}
\label{tab2}
\begin{tabular}{|l|c|c|c|}
\hline
      & $\chi^2$& dof & gof \\ \hline
Auger & 38.7    & 11  & $6\times 10^{-5}$\\
AGASA & 10.4    & 11  & 0.49\\
HiRes I+II & 16.5 & 19 & 0.62\\ \hline
\end{tabular}
\end{center}
\caption{Similar to Table~1, except that the results are for
 separate analyses of each experiment.
The analysis of the Auger dataset yields
a poor gof.}
\end{table}

\begin{figure}[ht]
\centering\leavevmode
\mbox{\includegraphics[width=6.5in]{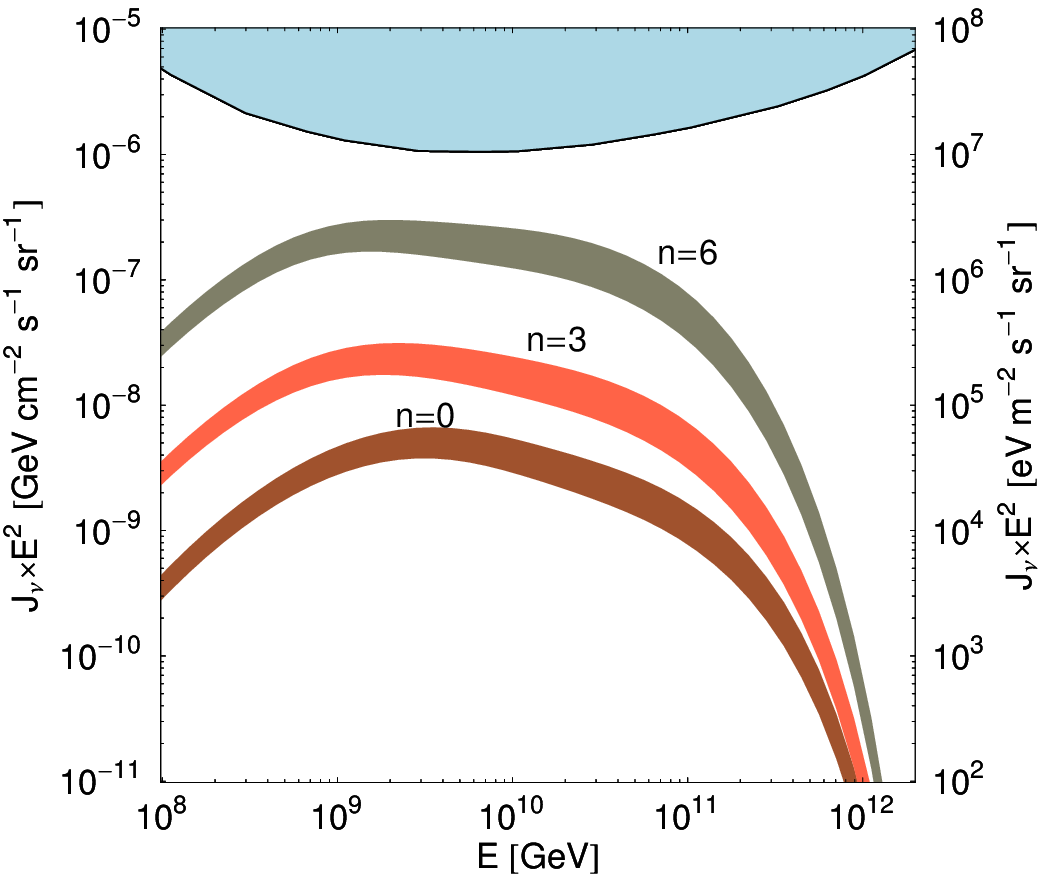}}
\caption[]{The  2$\sigma$ C.~L. determinations of the 
cosmogenic neutrino flux (summed over all flavors) 
corresponding to the injection spectra of 
Fig.~\ref{fig:protons}. The light shaded region is the
 95\% C.~L. model-independent
upper bound from RICE.
\label{fig:neut}}
\end{figure}

With the cosmic ray spectrum in hand we can compute the neutrino flux $J_\nu$ 
produced
in the GZK chain by using Eq.~(\ref{flux}) with $b=\nu$.
The resulting cosmogenic neutrino fluxes (summed over flavors) 
are shown in Fig.~\ref{fig:neut}. 
We do not show the case in which $n$ is free, since it fills the region
between the $n=0$ and $n=6$ bands.
Again, the bands 
correspond to the 2$\sigma$ C.~L. 
The 95\%~C.~L. model-independent
upper bound obtained by the RICE collaboration from their
latest data compilation is also shown.

We have seen that a combined analysis including Auger data has an unsatisfactory
goodness of fit even if super-GZK events seen by AGASA are excluded. This
suggests that an alternative method to obtain the injection spectrum
is called for.

\begin{figure}[ht]
\centering\leavevmode
\mbox{\includegraphics[width=6.5in]{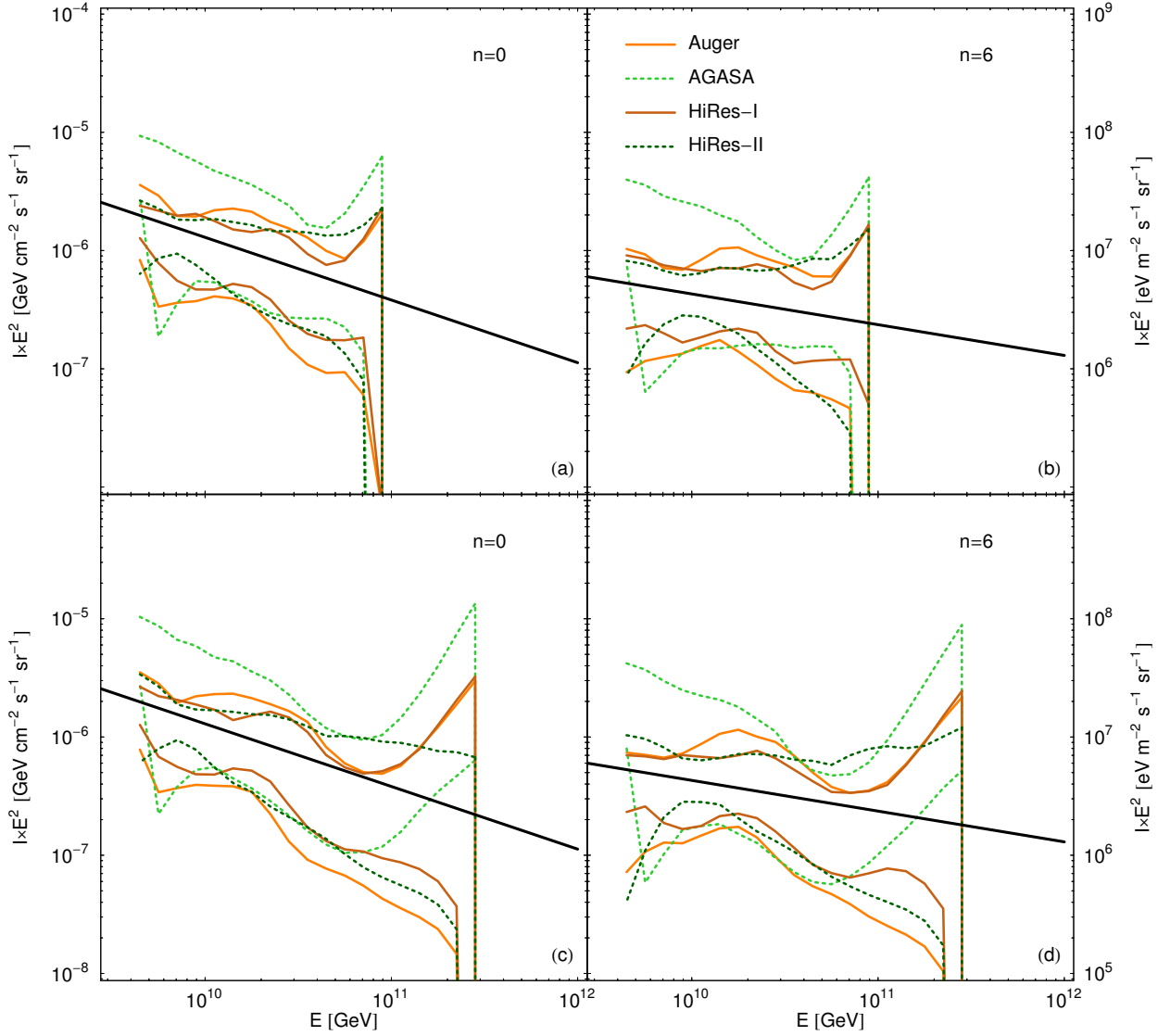}}
\caption[]{The 2$\sigma$ unfolded injection spectra from 
Auger, HiRes-I and II, and AGASA data.
The upper (lower) two panels are for an analysis of data with energy between 
$10^{9.6}$ GeV to $10^{11}$ GeV ($10^{9.6}$ GeV to $3\times 10^{11}$ GeV). 
Redshift evolution of the sources is parameterized by a power-law,
$(1+z)^n$. The bold line in the $n=0$ ($n=6$) 
plots is the injection
spectrum of the form $E_i^{-2.5}$ ($E_i^{-2.3}$). 

\label{fig:invprotons}}
\end{figure}

The injection spectra obtained by unfolding are 
shown in 
Fig.~\ref{fig:invprotons}. The widths of the bands are the 2$\sigma$ C.~L.
determinations. There are four different spectra since
the unfolding was carried out 
experiment-by-experiment; we have not attempted an inversion of the joint data
because there is too much freedom, rendering the results meaningless.
So that a straightforward comparison can be made
 with the power-law injection case, we have analysed 
data in the energy range $10^{9.6}$ GeV to $10^{11}$ GeV; see
panels (a) and (b)
of Fig.~\ref{fig:invprotons}. However,
since the unfolding method is applicable even for super-GZK events, we
have performed a separate analysis in which we include data up to 
$3\times 10^{11}$ GeV; see panels (c) and (d)
of Fig.~\ref{fig:invprotons}. 

In the unfolding procedure it is assumed that there is no flux beyond
the energy range spanned by the data. The sharp edges at the two ends of
the spectra result because the spectra can not be extrapolated outside the
energy interval. This is not true for the case in which a power-law injection 
spectrum is assumed.
The bold line corresponds to a power law
injection spectrum with spectral index $\alpha=-2.5$ in the $n=0$ plots and 
$\alpha=-2.3$ in the $n=6$ plots. The value of $\alpha$ is pinned-down by the
statistically significant data at the lower end of the energy range. Thus, we 
have used the same spectral index in panels (a) and (c) and in panels
(b) and (d).
Note that the bold lines are extrapolated outside the range covered
by the unfolding method. Results for values of $n$ between 0 and 6 are 
intermediate to those in the left-hand and right-hand panels of 
Fig.~\ref{fig:invprotons}.

\begin{figure}[ht]
\centering\leavevmode
\mbox{\includegraphics[width=6.5in]{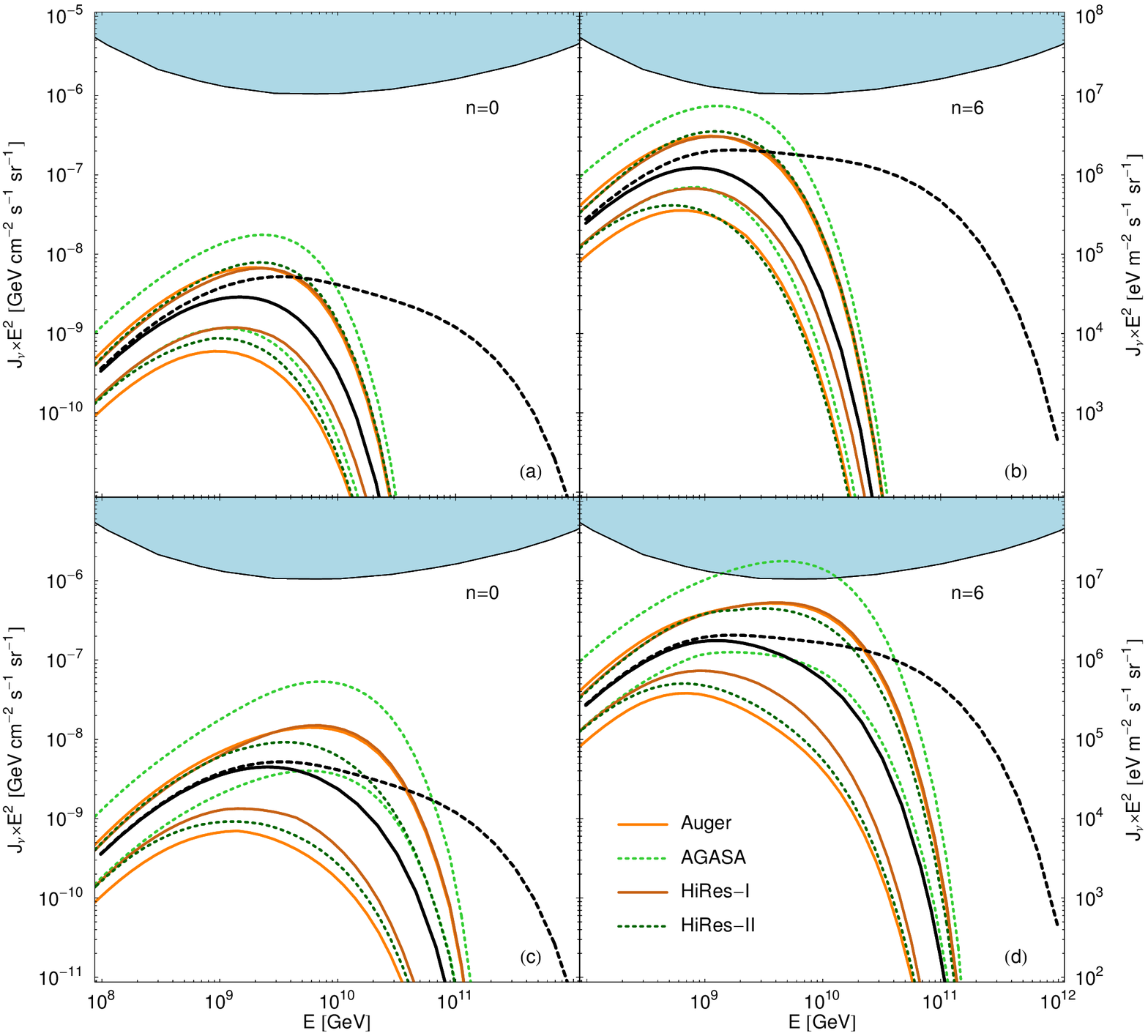}}
\caption[]{The  2$\sigma$ C.~L. determinations of the 
total cosmogenic neutrino flux
corresponding to the injection spectra of 
Fig.~\ref{fig:invprotons}. 
The bold solid lines in the upper and lower panels are for power-law injection 
spectra with 
\mbox{$E_{max}=10^{11}$ GeV} and $E_{max}=3\times 10^{11}$ GeV, respectively.
The agreement between the two methods is very good. 
The bold 
dashed lines are the neutrino spectra of Fig.~\ref{fig:neut}, which were
obtained for power-law injection spectra with $E_{max}=3\times 10^{12}$ GeV.
\label{fig:invneut}}
\end{figure}

The neutrino spectra in Fig.~\ref{fig:invneut} are obtained 
from Eq.~(\ref{neutinv}). The bands correspond to the 2$\sigma$ C~.L.
The bold solid lines in the top and bottom panels are for power-law injection 
spectra with 
$E_{max}=10^{11}$ GeV and $E_{max}=3\times 10^{11}$ GeV, respectively.
The bold dashed lines are the neutrino spectra of Fig.~\ref{fig:neut}  
and are provided for comparison. There we had set $E_{max}=3\times 10^{12}$ GeV.
 The agreement between the two methods is striking and
 provides support for the assumption that the injection spectrum has the
form of a power-law. The poor fit obtained in
 the combined analysis with a power-law spectrum is likely due
to systematic uncertainties that have not been accounted for.

Since the flux obtained from the unfolding procedure can be sizeable, 
 it is essential to check
that the accompanying cosmogenic photon flux 
is not in conflict with the EGRET
observation of the diffuse gamma ray flux~\cite{egret}, 
as emphasized in Ref.~\cite{semikoz}. 
We evaluated the photon flux using 
the publically available software, CRPropa~\cite{crpropa}{\footnote{Incidentally, 
we first confirmed that the cosmic ray spectra
generated using the propagation functions~\cite{prop2} and CRPropa 
are identical.}},
and find comfortable consistency with the EGRET bound. This is because
the injection spectra we have found fall steeply with energy and 
because the largest value of $E_{max}$ we consider is $3\times 10^{11}$ GeV.
As a consequence, the contribution to the photon flux measured by EGRET
is tiny.

\section{Constraints on the neutrino-nucleon cross section}

In the energy range of interest ($10^{8}$ to $10^{11}$ GeV) 
the strongest bounds on the neutrino flux
are those of the RICE experiment~\cite{rice}. (The current bound from the 
ANITA experiment is stronger than that from RICE only above
$10^{11}$ GeV~\cite{alite}).
The RICE bound is still
much weaker than the determinations we made in the
previous section. This allows us to place an upper 
bound on the neutrino-nucleon
cross section; new physics can not increase the Standard Model cross section
too much or high energy neutrinos would have been observed 
at neutrino telescopes.
Note that if the upper limit of the cosmogenic neutrino flux from either method
had been above the RICE limit, we could also have placed a constraint on the
minimum required suppression of the cross section at ultra high energies.

The RICE collaboration has employed a method to obtain 
neutrino flux limits that are independent of specific neutrino flux models. 
See Appendix II of Ref.~\cite{rice}. 
So long as new physics does not alter the energy dependence of the Standard
Model cross section drastically (and only changes the overall normalization),
 we can constrain the
neutrino-nucleon inelastic 
cross section by simply taking the ratio of the RICE bound to
that of the lower bound of the cosmogenic neutrino flux.

\begin{figure}[ht]
\centering\leavevmode
\mbox{\includegraphics[width=6.5in]{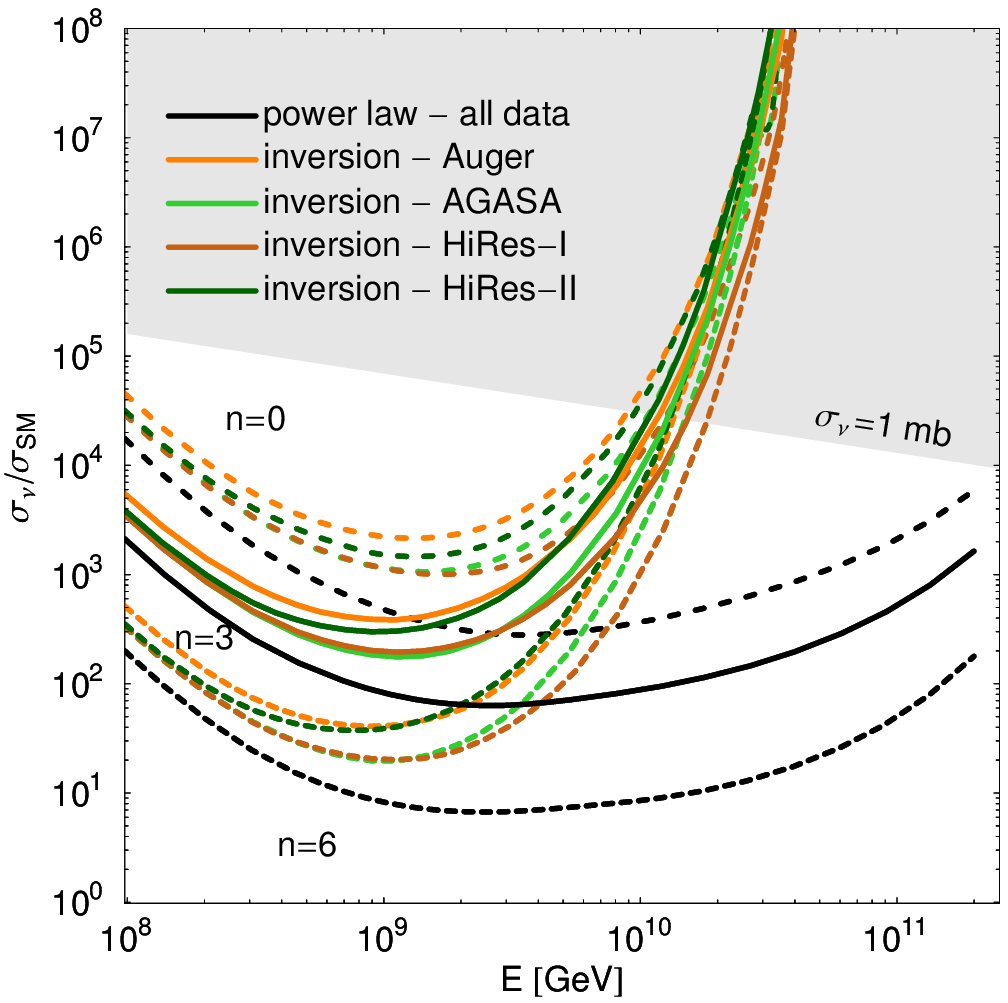}}
\caption[]{95\% C.L. upper bounds on the neutrino-nucleon cross section
using our determinations of the cosmogenic neutrino flux and the 
model-independent flux limit from RICE. Our bounds are only valid
in the unshaded region for which $\sigma_\nu<1$~mb.
\label{fig:cross}}
\end{figure}

Our 95\% C.L. upper bound on the cross section 
(in units of the Standard Model cross section) 
is shown in Fig.~\ref{fig:cross} for $n=0$, 3 and 6. 
In each case, the colored
curves correpond
 to the limit obtained by using the propagation inversion procedure
and the black curve corresponds to a power-law injection spectrum.
Bounds derived from RICE are
valid only for $\sigma_\nu \lsim 1$~mb~\cite{ring2}. Our cross section
bounds are applicable only in the unshaded region.

\section{Summary}

We obtained the proton injection spectrum from cosmic ray data using
two methods. In the first, we performed a combined analysis of 
HiRes-I and II, AGASA and Auger data in the energy interval, 
$10^{9.6}$ GeV to $10^{11}$ GeV, using a power law injection spectrum. 
We found the quality of the fit to be unsatisfactory even though we did not 
include super-GZK events in the analysis. The primary reason for the poor fit
is apparently that the initial Auger data are noisy. 
In the second method, we implemented
a regularized unfolding of the injection spectrum for each dataset separately.

We found the resultant cosmogenic neutrino flux using the injection spectra 
from both methods to be in excellent agreement with eachother. This suggests 
that the injection spectrum is well-modeled as a power-law and that the
poor fit mentioned above is due to unaccounted-for systematic uncertainties.

Using the model-independent limit on the neutrino flux from RICE, we constrained
the neutrino-nucleon cross section under the assumption that new physics 
modifies the charged-current and neutral-current interactions by the same 
constant factor. 

Our results are succintly summarized in Figs.~\ref{fig:protons}--\ref{fig:cross}.

\section{Acknowledgments}

We thank S.~Hussain and D.~W.~McKay for useful conversations and 
communications and D.~Bergman for providing us with the latest HiRes data.
We thank M.~Ahlers and A.~Ringwald for help
with the use of their propagation matrices. Furthermore we
acknowledge useful discussions with E.~Armengaud, T.~Beau and G.
Sigl on the CRPropa software.
This research was supported by the DOE
under Grant No.~DE-FG02-95ER40896, by the NSF
under CAREER Award No.~PHY-0544278 and Grant No.~EPS-0236913, 
by the State of Kansas through the
Kansas Technology Enterprise Corporation. 
Computations were performed on facilities 
supported by the
NSF under Grants No. EIA-032078 (GLOW), PHY-0516857
(CMS Research Program subcontract from UCLA), and PHY-0533280
(DISUN), and by the University of Wisconsin Graduate
School/Wisconsin Alumni Research Foundation.

\appendix

\section{$\chi^2$ analysis}

The observed flux $J_i^O$ is given by
$n_i^O/\epsilon_i$, where $n_i^O$ and $\epsilon_i$ is the binned event rate
and effective exposure, respectively, in bin $i$ of an experiment.
 The theoretically predicted flux $J_i^t$ is
obtained from Eqs.~(\ref{flux},\ref{eq:powerlaw}). 
It depends on the overall
normalization  $\rho_0I_0$, the spectral index $\alpha$ and on
redshift evolution as parameterized by $n$. 
The theoretical event rate $n_i^t$ is related to the
theoretical flux by $n_i^t=J_i^t\epsilon_i$. 

In order to incorporate the effect of the energy scale uncertainty 
on the event rates, we replace the
(observed) energy $E_i$ by $(1+\gamma)E_i$, where $\gamma$ is the
fractional energy scale uncertainty of the experiment.  
The details follow the
implementation in GLoBES~\cite{Huber:2004ka}.
Thus, the theoretical flux becomes
a function of $\rho_0I_0, \alpha ,n$ and 3 $\gamma$'s.

Since the the observed event rate can be very low or even zero, we
 use the Poissonian $\chi^2$-function (see, {\it e.g.}~\cite{PDB}),
\begin{equation}
\label{eq:poiss}
\chi^2(n_i^O,n
_i^t)=2\sum_{i=1}^{N}\left(n_i^t-n_i^O+n_i^O\log\frac{n_i^O}{n_i^t}\right)\,.
\end{equation}
It is crucial to propagate the
errors and their correlations consistently as we are interested in both, 
the proton flux and
the derived cosmogenic neutrino flux. The supposition that the errors are
uncorrelated would lead to an incorrect result. The allowed region at
confidence level $CL$ is defined by requiring that
$\Delta\chi^2\leq\chi^2_{CL}$, where $\chi^2_{CL}$ is given by the
$CL^{th}$ percentile of the $\chi^2$-distribution after accounting for the
 5 or 6 free parameters in the analysis\footnote{3 energy scales, 1 normalization, 1 spectral index and
  optionally the evolution parameter $n$.}. 
We use a Markov Chain Monte Carlo (MCMC) method~\cite{MCMCintro} based on the
Metropolis-Hastings algorithm. 
Specific care has to be taken to ensure that the chain has
reached equilibrium, {\ie} its asymptotic state. We 
employ the convergence diagnostic
of Ref.~\cite{Dunkley:2004sv} since it is conceptionally
simple and does not incur a large computational burden. 

\section{Unfolding procedure}

\begin{figure}[ht]
\centering\leavevmode
\mbox{\includegraphics[width=6.5in]{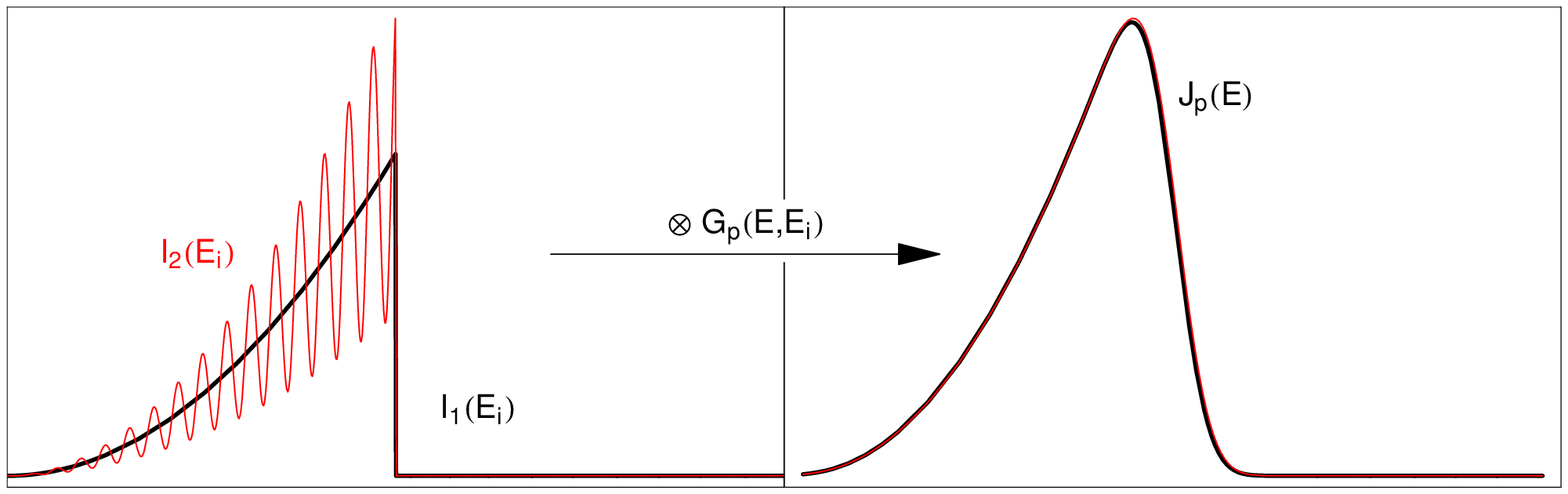}}
\caption[]{Two different injection spectra, $I_1(E_i)$ and $I_2(E_i)$, 
that yield the
same cosmic ray spectrum $J_p(E)$ on folding in $G_p(E,E_i)$. 
A naive unfolding  
of data corresponding to $J_p(E)$ ({\it{i.e.,}} 
with random fluctuations included) 
could result in a highly oscillatory injection spectrum like $I_2(E_i)$ 
which may be not be positive definite at all energies. The goal of regularized
unfolding is to extract the smooth spectrum $I_1(E_i)$ with the minimal
introduction of bias.
\label{illposed}}
\end{figure}

The propagation of ultra high energy cosmic rays 
is described by Eq.~(\ref{flux}) which is a
 Fredholm equation of the first kind.
The authors of Ref.~\cite{ring} have attempted 
to infer $I(E_i)$ from the observed flux $J_b(E)$ and $G_b(E,E_i)$
  via direct inversion.
Unfortunately, the problem is ill posed unless regularized unfolding is
carried out; see Ref.~\cite{Craig} for a detailed exposition. An illustration
of this fact is provided in Fig.~\ref{illposed}.

We follow Tikhonov regularization described in  Ref.~\cite{unfolding}. 
The $\chi^2$ definition of Eq.~(\ref{eq:poiss}) must be modified 
to include a
regularization function 
$S(\mathbf{n}^t)$ and a regularization parameter $\beta$,
\begin{equation}
\label{eq:reg_chi}
\chi^2(\mathbf{n}^t)+\beta S(\mathbf{n}^t)\,,
\end{equation}
with the understanding that the $n^t_i$ are the parameters in the fit;
\ie\ the $n^t_i$ are not derived from a model. It
can be shown that $\chi^2$-minimization with $\beta=0$ is an 
unbiased estimator with the smallest variance but which yields highly
oscillatory solutions.
One the other hand, 
the $\beta\rightarrow\infty$ case yields a maximally smooth estimator
 with vanishing
variance and a clear bias (since the result no longer depends on the data). 
The parameter $\beta$ determines the relative weight placed on the data
in comparison to the degree of smoothness of the solution. 
In a Bayesian spirit this is equivalent to assuming a prior which
favors smooth solutions with $\beta$ controlling the width of the
prior. The smaller $\beta$ is, the less impact the prior will
have and vice versa.

It remains to choose $S(\mathbf{n}^t)$ (to obtain a smooth solution)
 and $\beta$ (to define the trade-off between bias and variance).
For our application, the mean square of the second derivative
\begin{equation}
S(\mathbf{\mathbf{n}^t})=-\sum_{i=1}^{N-2}\left(-n_i^t+2n_{i+1}^t-n_{i+2}^t\right)^2\,,
\end{equation}
works very well. 

Our criteria for obtaining $\beta$ are that injection spectrum be
positive definite and that the overall $\chi^2$ for an individual dataset 
not increase by more
than 50\% relative to that obtained from the power-law fit. 
Values of $\beta$ above unity satisfy the former requirement, while
$\beta$ needs to be less than 10 to satisfy the latter.
We select
$\beta=5$ for all four datasets, 
which adequately suppresses any spurious oscillations. 

We estimate the variance of the result by generating 1000 random
realizations of the data sets by assuming a Poisson distribution in
each bin with a mean value $n_i^O$. For each of these random
realizations we repeat the above procedure and construct the
covariance matrix $\mathbf{C}$ which is then used to obtain upper
and lower bounds on the proton and neutrino fluxes by retaining only
those realizations which satisfy
\begin{equation}
\mathbf{n}^t\mathbf{C}^{-1}(\mathbf{n}^t)^T\leq\chi^2_{CL}\,.
\end{equation}
Here, $\chi^2_{CL}$ is given by the $CL^{th}$ percentile of the values
obtained for the left hand side.


\end{document}